\documentclass{osa-article}

\journal{oe}
\usepackage{color}

\articletype{Research Article}

\begin{document}

\title{Generation of subnatural-linewidth
 orbital angular momentum entangled biphotons using a single driving laser in hot atoms}

\author{Jiaheng Ma,\authormark{1,\dag} Chengyuan Wang,\authormark{1,\dag,3} Bingbing Li,\authormark{1} Yun Chen,\authormark{2} Ye Yang,\authormark{1} Jinwen Wang,\authormark{1}  Xin Yang,\authormark{1} Shuwei Qiu,\authormark{1} Hong Gao,\authormark{1,4} and Fuli Li\authormark{1}}

\address{\authormark{1}MOE Key Laboratory for Nonequilibrium Synthesis and Modulation of Condensed Matter, Shaanxi Province Key Laboratory of Quantum Information and Quantum Optoelectronic Devices, School of Physics, Xi`an Jiaotong University, 710049, China\\
\authormark{2}Department of Physics, Huzhou University, Huzhou 313000, China\\
\authormark{\dag}These authors contributed equally to this work
}
\email{\authormark{3}wcy199202@gmail.com} 
\email{\authormark{4}honggao@xjtu.edu.cn}



\begin{abstract}
Orbital angular momentum (OAM) entangled photon pairs with narrow bandwidths play a crucial role in the interaction of light and quantum states of matter. In this article, we demonstrate an approach for generating OAM entangled photon pairs with a narrow bandwidth by using a single driving beam in a $^{85}$Rb atomic vapor cell. This single driving beam is able to simultaneously couple two atomic transitions and directly generate OAM entangled biphotons by leveraging the OAM conservation law through the spontaneous four-wave mixing (SFWM) process. The photon pairs exhibit a maximum cross-correlation function value of 27.7 and a linewidth of 4 MHz. The OAM entanglement is confirmed through quantum state tomography, revealing a fidelity of 95.7\% and a concurrence of 0.926 when compared to the maximally entangled state. Our scheme is notably simpler than previously proposed schemes and represents the first demonstration of generating subnatural-linewidth entangled photon pairs in hot atomic systems.
\end{abstract}

\section{Introduction}
Entangled photon pairs, also known as entangled biphotons, are fundamental quantum resources with widespread applications in fields such as quantum communication, quantum computing, and quantum cryptography\cite{01,02,03,04}. Multiple degrees of freedom (DoFs) of photons, such as polarization\cite{02,wang2020generation,park2021direct}, path\cite{05}, time-energy (or frequency)\cite{06,kim2022quantum}, position-momentum\cite{wang2019generation}, and orbital angular momentum (OAM)\cite{07,PhysRevLett.116.073601,PhysRevLett.124.083605}, can be harnessed for entanglement generation. Among these DoFs, OAM can be utilized to construct a complete infinite-dimensional Hilbert space, thereby facilitating the generation of high-dimensional entanglement states to significantly boost quantum information capacity\cite{erhard2020advances,08,09}.

The initial generation of OAM entangled photon pairs was accomplished through the process of spontaneous parametric down-conversion (SPDC) in nonlinear crystals\cite{07}. Subsequently, OAM entangled photon pairs have also been successfully prepared using the mechanisms of spontaneous Raman scattering (SRS) and spontaneous four-wave mixing (SFWM) in atomic ensembles\cite{10,11}. The entangled biphotons generated by SPDC typically have a large bandwidth, which is unsuitable for atomic-based quantum information processing\cite{01,09,wang2021efficient,shen2023shape}. Introducing an optical cavity or waveguide\cite{12,13} can mitigate this issue, but it increases the complexity of the experimental setup. Biphotons generated by the SFWM process in cold atoms exhibit a subnatural linewidth with the assistance of the Electromagnetically Induced Transparency (EIT) effect\cite{14,15}. However, constructing a cold atom system requires the implementation of a sophisticated timing control system and a complex magneto-optical trap (MOT) system\cite{16}. Conversely, hot atomic ensembles offer a relatively simpler and more cost-effective alternative, yet it also comes with certain limitations that need to be addressed.

In a hot atomic ensemble, the frequent collisions between atoms and the container walls result in rapid atomic repopulation between different ground states, which leads to uncorrelated photon noise. Besides, the Doppler broadening caused by atomic thermal motion will bring about strong resonance fluorescence noise. These drawbacks directly impact the production rate, signal-to-noise ratio, and bandwidth of biphotons\cite{18,19}. Employing paraffin-coated cells and introducing an optical pumping beam can help to eliminate atomic repopulation-induced noise and prolong the biphoton coherence time\cite{20,21,PhysRevResearch.4.023132}. Increasing the detuning of the pump beam can separate the frequency of the signal photons from the resonance fluorescence noise\cite{20,wang2018efficient,davidson2021bright}. In the ladder-type and double $\Lambda$-type SFWM schemes, the frequency of the experimental beams is similar to that of the biphotons, which is difficult to filter through narrow-band filters and gives rise to considerable noise. This kind of noise in a cold atomic ensemble is generally suppressed by introducing a small separation angle between the experimental beams and biphotons\cite{22,15,24}. Nevertheless, this approach will introduce residual phase mismatch in SFWM, and the Doppler shift in the hot atomic ensemble will destroy the EIT condition and limit the biphoton coherence time\cite{25,26,27}. Recently, Yu et al. proposed an all-collinear propagation scheme, wherein the experimental beams and the photon pairs propagate in the same direction to satisfy the phase-match condition\cite{28}. This configuration allows for the generation of biphotons with sub-megahertz linewidth and high spectral brightness.

Recent studies have shown the production of OAM entangled biphotons in hot atoms \cite{wang2020generation,Shi:20}; however, the bandwidth of these biphotons remains too large to be efficiently utilized in atomic-based quantum memory\cite{29,30}, quantum wavelength conversion\cite{31}, and quantum phase gates\cite{32}, etc. Moreover, current approaches for the generation of OAM entangled photon pairs necessitate at least two frequency-stabilized narrow linewidth lasers as the experiment beams, as well as some other beams for optical system alignment, which makes the system spatially bulky and resource consuming\cite{33,34}. 

In this work, we present a simple scheme for directly generating OAM entangled photon pairs in a hot atomic vapor cell using only one driving laser. This laser can simultaneously couple two atomic transitions to form a double $\Lambda$-type SFWM, resulting in the generation of paired Stokes photon and anti-Stokes photon. The maximum cross-correlation function value of the biphoton is measured to be 27.7, violating the Cauchy–Schwarz inequality by a factor of 191.8. Additionally, the all-collinear structure enables the driving beam to create an EIT transparent window for the anti-Stokes photon, which prolongs the biphoton’s coherence time to 40 ns (corresponding to a bandwidth of 4 MHz). We further verify the OAM entanglement by quantum state tomography and thereby demonstrate a fidelity of 95.7\% and concurrence of 0.926 compared with the maximally entangled state.

\section{Experimental configuration}
\begin{figure*}[!htbp]
  \centering
  \includegraphics[width=5.0in]{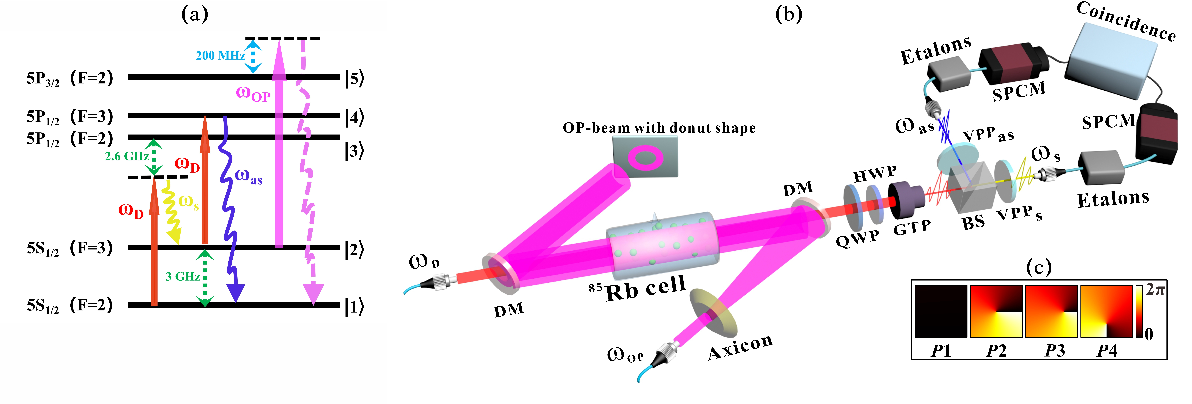}
\caption{Generation of OAM entangled photon pairs using a single driving beam. (a) The energy diagram of SFWM, (b) the schematic diagram of the experimental setup, and (c) phase distribution maps corresponding to the four modes of the VPP. (DM, dichroic mirrors; QWP, quarter-wave plate; HWP, half-wave plate; GTP, Gran-Taylor prism; BS, beam splitter; VPP, vortex phase plate; SPCM, single-photon counting module.)}
\label{fig:1}
\end{figure*}

The energy diagram and experimental setup are schematically illustrated in Fig. \ref{fig:1}(a) and \ref{fig:1}(b). A single driving beam ($\rm{\omega_{D}}$, 795 nm) with horizontal polarisation and a diameter of 3 mm acts as both the pump beam and the coupling beam in SFWM. This beam first couples the transition of $\left|5 \mathrm{S}_{1 / 2}, \mathrm{F}=2\right\rangle \rightarrow\left|5 \mathrm{P}_{1 / 2}, \mathrm{F}=2\right\rangle$ with 2.64 GHz red detuning and stimulates a Stokes photon. Then it resonatly couples the transition of $\left|5 \mathrm{S}_{1 / 2}, \mathrm{F}=3\right\rangle \rightarrow\left|5 \mathrm{P}_{1 / 2}, \mathrm{F}=3\right\rangle$ to stimulates an anti-Stokes photon. Each pair of Stokes photon and anti-Stokes photon is generated almost simultaneously by the atomic ensemble. Hence the photon pairs follow the OAM conservation law and are naturally in OAM entanglement\cite{offer2018spiral,07}. An additional horizontally polarized optical pumping beam with ($\rm{\omega_{OP}}$, 780 nm), which is blue detuned from $\left|5 \mathrm{S}_{1 / 2}, \mathrm{F}=3\right\rangle$ to $\left|5 \mathrm{P}_{3 / 2}, \mathrm{F}=2\right\rangle$ by 200 MHz, can effectively populate most atoms across all Zeeman sub-levels of $\left|1 \right\rangle$ to counteract atomic repopulation-induced noise. In order not to interfere with the SFWM transitions, this beam is modulated into a donut shape with its inner radius larger than the diameter of the driving beam such that there is no spatial overlap between them \cite{21}. As shown in Fig. \ref{fig:1}(b), the driving beam and the optical pumping beam counter-propagate in a $^{85} \mathrm{Rb}$ atomic cell, which is coated with paraffin inside and has a length of 80 mm. The donut shaped optical pumping beam (with an inner radius of 5 mm and an outer radius of 12 mm) is obtained by using an axicon and a lens and is reflected into the cell through a dichroic mirror (DM).

The horizontally polarized driving beam could stimulate the photon pairs that are both horizontally polarized or vertically polarized. The primary issue to be addressed in collinear structure is to isolate the biphotons from the driving beam, which has a much higher intensity (the photon number of the driving beam is $10^{10}$ times larger than that of the biphoton). To achieve this, we utilize two Gran-Taylor prisms (GTPs) in series as polarization filters, followed by three etalons in series located in the Stokes and anti-Stokes paths as frequency filters. The GTPs can filter the vertically polarized biphotons from the horizontally polarized driving beam by an extinction ratio of 85 dB. The center frequency of the etalon filters, with a transmission bandwidth of 500 MHz, matches that of the Stokes (anti-Stokes) photons in the corresponding optical path. This enables the isolation of noise photons with frequencies differing from the matched center frequency with an extinction ratio of 75 dB. After these filter elements, the photon number of the driving beam can be decreased to a hundred per second, which is negligible compared with the biphoton's generation rate. A beam splitter (BS) separates Stokes photons (transmitted) and the anti-Stokes photons (reflected) into two paths and the photons are subsequently directed into Single Photon Counting Modules (SPCMs) after being collected by single-mode fibers (SMFs). Considering the transmittance of all devices and the detection efficiency of the SPCM, the overall collection efficiency of the Stokes photons (or anti-Stokes photons) is measured to be 4.0\% (3.2\%).

By using this experimental device, we have successfully prepared biphotons. The SFWM process exhibits phase coherent, both longitudinally and transversely, between the driving beam and the generated biphotons\cite{offer2018spiral}. This feature enables the OAM conservation and inherently induces OAM correlations between the biphotons\cite{offer2018spiral,07}. The system's initial condition of zero linear and angular momentum leads to biphotons possessing zero total OAM. In this case, the representation for the biphoton entanglement should be \cite{07} 
\begin{equation}
|\Psi\rangle=\sum_{l=-\infty}^{l=\infty}C_l({\left|l\right\rangle}_s\otimes{\left|-l\right\rangle}_{as}),
\label{eq(1)}
\end{equation}
where $\left|l\right\rangle$ denotes the OAM state of the $l$ quantum eigenmode and $C_l$ is the coefficient. Considering only two-dimensional OAM entanglement, the photon pairs will be in the state of
\begin{equation}
|\Psi\rangle=\frac{1}{\sqrt2}(\left|G_sG_{as}\right\rangle+\left|R_sL_{as}\right\rangle),
\label{eq(2)}
\end{equation}
where $G$ , $R$, and $L$ denote the OAM states with topological charge $l$ = 0, 1, and -1, respectively. Taking into account the phase flip of the BS to the photons in the reflection direction, we rewrite the above formula in the detector coordinate system as
\begin{equation}
    |\Psi\rangle=\frac{1}{\sqrt2}                   (\left|G_sG_{as}\right\rangle+\left|R_sR_{as}\right\rangle).
    \label{eq(3)}
\end{equation}

The vortex phase plate (VPP) serves the purpose of converting a Gaussian mode beam into an OAM beam and vice versa. This allows us to utilize the VPPs and SMFs for performing projection measurements on photon pairs. Different regions of the VPP correspond to different measurement bases: the edge region of the VPP, characterized by an approximately uniform phase distribution, is suitable for measuring the $\left|G \right\rangle$ state ($P1$ mode); the central region can measure the $\left|R \right\rangle$ state ($P2$ mode); by shifting the phase plate, we can measure the \((\left|G \right\rangle+\left|R \right\rangle)/\sqrt{2}\) state ($P3$ mode); rotating the $P3$ mode by 90° enables us to perform measurements on the \((\left|G \right\rangle-i\left|R \right\rangle)/\sqrt{2}\) state ($P4$ mode). The phase distribution diagrams corresponding to the four modes of the VPP are depicted in Fig. \ref{fig:1}(c). Through projecting photons in these four bases, a total of 16 sets of coincidence count values can be obtained, which are denoted as: $\left|P1_s,P1_{as} \right\rangle$, $\left|P1_s,P2_{as} \right\rangle$, $\left|P1_s,P3_{as} \right\rangle$, $\left|P1_s,P4_{as} \right\rangle$, $\left|P2_s,P1_{as} \right\rangle$, $\left|P2_s,P2_{as} \right\rangle$, $\left|P2_s,P3_{as} \right\rangle$, $\left|P2_s,P4_{as} \right\rangle$, $\left|P3_s,P1_{as} \right\rangle$, $\left|P3_s,P2_{as} \right\rangle$, $\left|P3_s,P3_{as} \right\rangle$, $\left|P3_s,P4_{as} \right\rangle$, $\left|P4_s,P1_{as} \right\rangle$, $\left|P4_s,P2_{as} \right\rangle$, $\left|P4_s,P3_{as} \right\rangle$, and $\left|P4_s,P4_{as} \right\rangle$, respectively. By utilizing these 16 sets of data for quantum state tomography, we can reconstruct the density matrix of our photon pairs.
\section{Experimental results and discussions}

Firstly, we investigate the coincidence count rate and the cross-correlation function $\rm{g^2_{s,as}(\tau)}$ of the photon pairs as functions of cell temperature ($\rm{T_{cell}}$), optical pumping beam power ($\rm{P_{OP}}$), and driving beam power ($\rm{P_D}$). Here, $\rm{g^2_{s,as}(\tau)}$ is defined as
\begin{equation}
    \mathrm{g^2_{s,as}(\tau)=\frac{\langle{n_s(t)\cdot{n_{as}(t+\tau)}}\rangle}{\langle{n_s(t)}\rangle\langle{n_{as}(t)}\rangle}},
    \label{eq(4)}
\end{equation}
where $\tau$ is the time interval between the photon pairs, $\mathrm{n(t)}$ is the photon number detected by the SPCM at time $\mathrm{t}$, and $\mathrm{\langle{n}\rangle}$ is to take the average. In the following, we extract the maximum cross correlation function value $[\rm{g^2_{s,as}]_{m}}$ from $\rm{g^2_{s,as}(\tau)}$ for data analysis.

\begin{figure*}[!htbp]
    \centering    \includegraphics[width=5.0in]{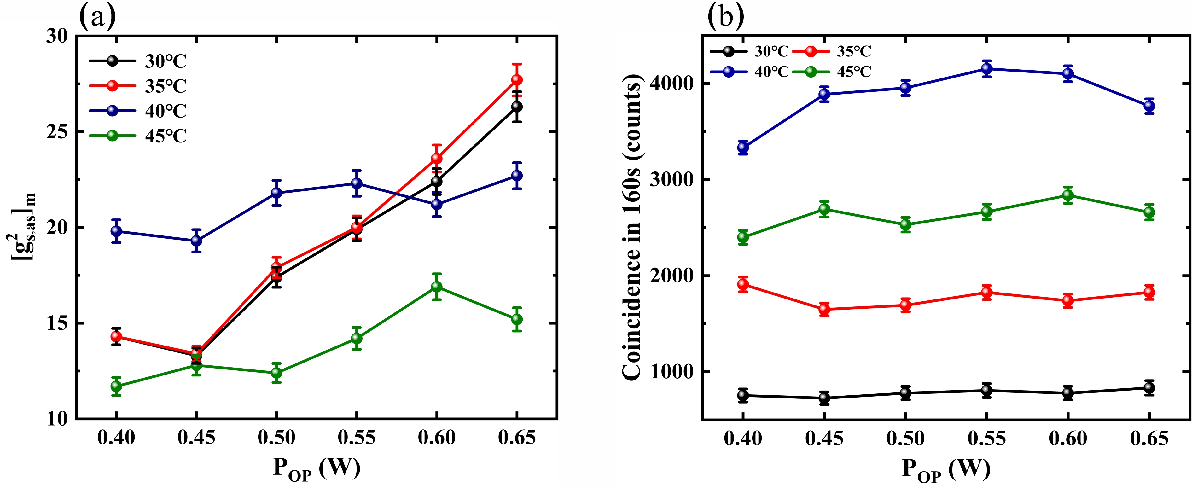}
    \caption{The variation of  $\rm{[g^2_{s,as}]_m}$ (a) and coincidence (b) with $\rm{P_{OP}}$ and $\rm{T_{cell}}$. The photon pairs are in the  $\left|R_sR_{as} \right\rangle$ state. $\rm{P_D}$ remains constant at 38 mW, and the accumulation time is 160 seconds.}
    \label{fig:2}    
\end{figure*}

We begin by measuring $[\rm{g^2_{s,as}]_{m}}$ and the coincidence counts (in 160 s) of photon pairs in the $\left|R_sR_{as} \right\rangle$ state with respect to $\rm{P_{OP}}$ under different $\rm{T_{cell}}$, as depicted in Fig. \ref{fig:2}. Considering that the paraffin is destroyed at high temperatures, $\rm{T_{cell}}$ is varied from 30 ℃ to a maximum of 45 ℃. We maintained $\rm{P_D}$ at the maximum achievable of 38 mW. Under low $\rm{T_{cell}}$ (30 ℃ and 35 ℃), $[\rm{g^2_{s,as}]_{m}}$ increases with growing $\rm{P_{OP}}$. However, as $\rm{T_{cell}}$ reaches 45 ℃, the rate of increase in $[\rm{g^2_{s,as}]_{m}}$ slows down and starts to descend after 0.6 W of $\rm{P_{OP}}$. The coincidence rate is not significantly affected by $\rm{P_{OP}}$ at each $\rm{T_{cell}}$. These phenomena can be explained as follows. The pairing ratio of the biphotons increases with higher atomic density ($\rm{T_{cell}}$), resulting in a larger $[\rm{g^2_{s,as}]_{m}}$. But excessive atomic density will lead to the re-absorption of anti-Stokes photons and increase uncorrelated noise. One main reason for the increased uncorrelated noise is that atoms initially populated on $\left|2\right\rangle$ are excited by the coupling beam to stimulate noise photons that have the same frequency as the anti-Stokes photons. Therefore, the OP beam is applied to pump atoms from $\left|2\right\rangle$ to $\left|1\right\rangle$. To avoid interference with the SFWM transitions, this beam is transformed into a donut shape, with its inner radius larger than the diameter of the driving beam, thereby ensuring spatial separation between them. However, a small fraction of the OP beam is dispersed into the hollow region due to diffraction. Under a large $\rm{T_{cell}}$ and $\rm{P_{OP}}$, atoms oscillate more frequently between the OP region and the SFWM region, enhancing the likelihood of atoms returning directly to $\left|1\right\rangle$ via the OP beam instead of stimulating anti-Stokes photons. This leads to an incomplete pairing of Stokes and anti-Stokes photons and a decreased $[\rm{g^2_{s,as}]_{m}}$.

\begin{figure*}[b]
    \centering
    \includegraphics[width=5.0in]{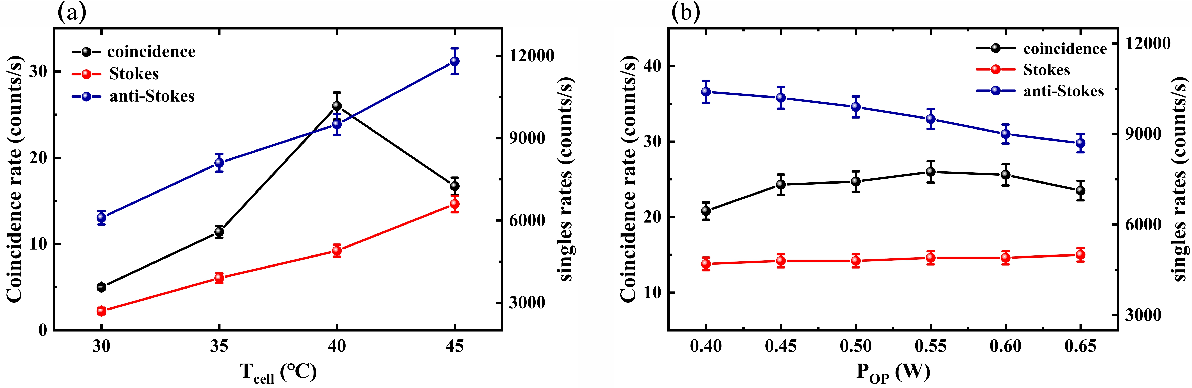}
    \caption{The relationship between the variation trends of single path generation rate and coincidence count rate. Photon pairs are in the $\left|R_sR_{as} \right\rangle$ state and $\rm{P_D}$ remains constant at 38 mW. The $\rm{P_{OP}}$ is set to 0.55 W, with the
$\rm{T_{cell}}$ range of 30 to 45 ℃, (a). The $\rm{T_{cell}}$ is set to 40 ℃, while $\rm{P_{OP}}$ varying between 0.40 and 0.65 W, (b). }
    \label{fig:3}   
\end{figure*}

We further investigate the relationship between the variation in the generation rate of Stokes and anti-Stokes photons and the change in coincidence count rate. The photon pairs remain in the $\left|R_sR_{as} \right\rangle$ state, and experimental results are illustrated in Fig. \ref{fig:3}. Based on prior experimental results, we vary $\rm{T_{cell}}$ from 30 to 45 °C, with a fixed $\rm{P_{OP}}$ of 0.55 W. Additionally, at $\rm{T_{cell}}$ of 40 °C, we vary $\rm{P_{OP}}$ from 0.40 to 0.65 W. As depicted in Fig. \ref{fig:3}(a), with an increase in $\rm{T_{cell}}$, both the generation rate of Stokes photons and anti-Stokes photons rise. However, the coincidence rate reaches its peak at $\rm{T_{cell}}$ of 40 °C and then begins to decline due to the influence of thermal photon noise. As $\rm{P_{OP}}$ increases, the uncorrelated photons from the on-resonance Raman scattering of the coupling beam are suppressed, resulting in a reduction in the anti-Stokes photon count rate and a growth of the coincidence count rate. However, if $\rm{P_{OP}}$ is too large (above 0.6 W), the possibility of atoms directly returning to $\left|1\right\rangle$ rather than emitting anti-Stokes photons will be increased, leading to a decline in the coincidence rate, as shown in Fig. \ref{fig:3}(b).

\begin{figure*}[t]
     \centering
     \includegraphics[width=5.0in]{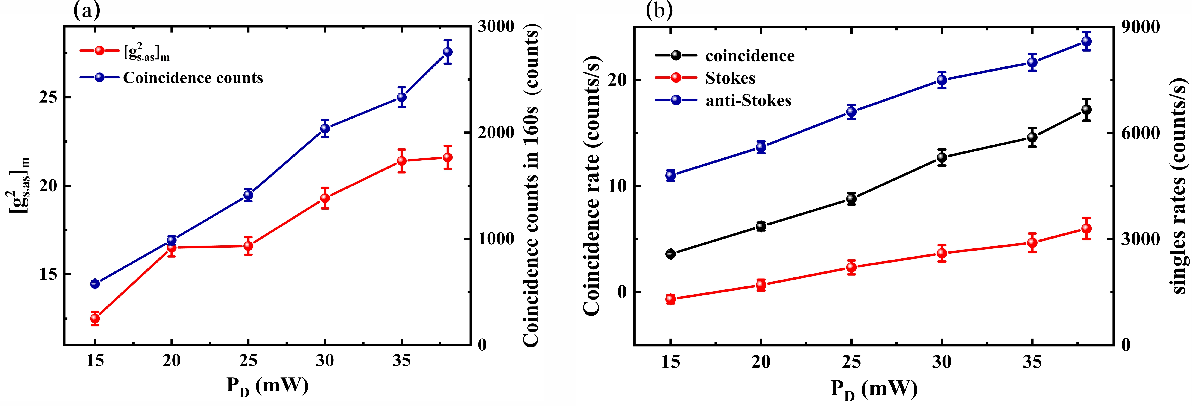}
    \caption{The variation trends of coincidence count and  $\rm{[g^2_{s,as}]_m}$ as $\rm{P_D}$, (a). The variation trends of coincidence count rate and singles rate as $\rm{P_D}$, (b). Photon pairs are in the $\left|R_sR_{as} \right\rangle$ state. The accumulation time is set to 160 seconds and $\rm{P_D}$ is varied from 15 to 38 mW. $\rm{T_{cell}}$ is set to 40 ℃, while $\rm{P_{OP}}$ is set to 0.55 W.}
        \label{fig:4}
\end{figure*}
After analysing the influences of $\rm{T_{cell}}$ and $\rm{P_{OP}}$, we investigated the variation trends of coincidence count rate, $[\rm{g^2_{s,as}]_{m}}$, and single path generation rate as $\rm{P_D}$ varied from 15 to 38 mW. Based on previous research findings, we set the time-resolved coincidence parameters for photon pairs in the $\left|R_sR_{as} \right\rangle$ state as $\rm{T_{cell}}$ 35 ℃ and $\rm{P_{OP}}$ 0.60 W. The experimental results are shown in Fig. \ref{fig:4}. Fig. \ref{fig:4}(a) illustrates that as $\rm{P_D}$ increases, both $\rm{g^2_{s,as}(\tau)}$ and coincidence rate of photon pairs exhibit an upward trend. Additionally, Fig. \ref{fig:4}(b) reveals that the generation rates of Stokes photons and anti-Stokes photons in the $\left|R_sR_{as} \right\rangle$ state also rise as $\rm{P_D}$ increases.

\begin{figure*}[!htbp]
     \centering
     \includegraphics[width=5.0in]{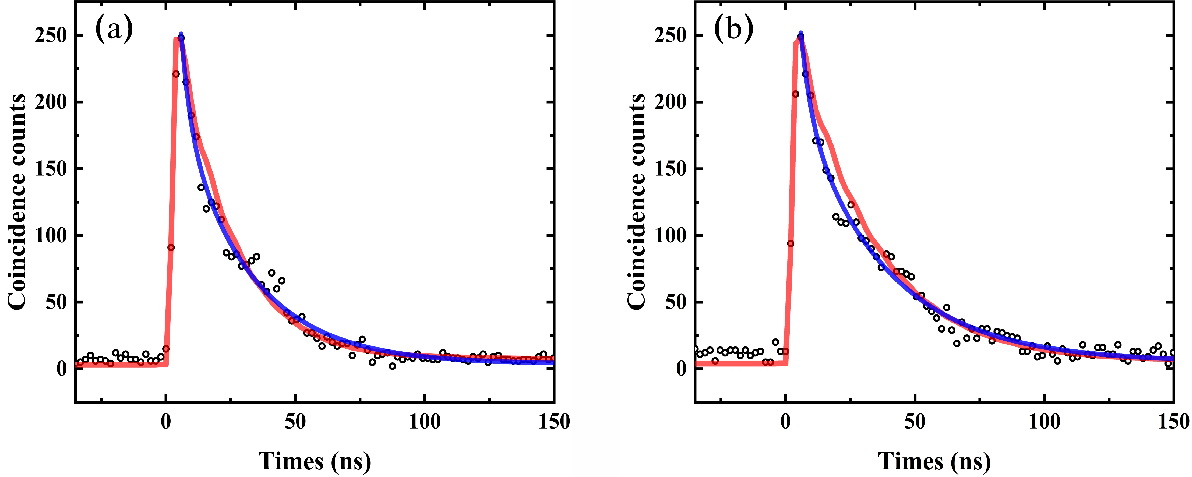}
      \caption{The time-resolved coincidence results of photon pairs in the $\left|G_sG_{as} \right\rangle$ state (a) and $\left|R_sR_{as} \right\rangle$ state (b). The accumulation time is 160 s, and the time bin is 1.94 ns. The black dots represent the number of coincidences per time bin, the red line represents the theoretical prediction, and the blue line represents the fitted curve for the decaying coincidences. The time-resolved coincidence parameters are $\rm{T_{cell}}$ $=$ 40 ℃ and $\rm{P_{OP}}$ $=$ 0.55 W. }
            \label{fig:5}
\end{figure*}
The optimal measurement parameters for the $\left|G_sG_{as} \right\rangle$ state photon pairs are essentially the same as those for the $\left|R_sR_{as} \right\rangle$ state photon pairs. Therefore, we set the $\rm{P_D}$ to its maximum value of 38 mW, which is the upper limit of our laser. The other parameters are $\rm{T_{cell}}$ set at 40 ℃ and $\rm{P_{OP}}$ at 0.55 W. The time-resolved coincidence results of these two states with an accumulating time of 160 seconds are shown in Fig. \ref{fig:5} with black dots. The red lines represent the theoretical predictions based on the theoretical model\cite{28}. We also employ an exponential function, e.g., $y(x)=y_{0}+A\exp{\left[-x/\tau\right]}$, to fit the decay behavior of the biphoton wave packets, which is depicted by the blue lines in the figure. We derive the baseline value $y_0$ based on the experimental data, and employ the fitting parameters, namely the amplitude $A$ and time constant $\tau$, to attain the optimal fitting curve. The 1/e coherence time $\tau _{co}$ derived from the fitting curve is approximately 40 ns, and the biphoton linewidth is determined as $1 / (2\pi \tau _{co})$ = 4 MHz. This extended coherence time can be attributed to the all-collinear propagation structure, which establishes an EIT transparent window for the anti-Stokes photons, effectively prolonging the coherence time of the photon pairs. Additionally, in Fig. \ref{fig:5}(a) and \ref{fig:5}(b), the maximum coincidence count values and background noise levels are nearly identical. This similarity suggests that the collection efficiency for both $\left|G_sG_{as} \right\rangle$ and $\left|R_sR_{as} \right\rangle$ state photon pairs is consistent, indirectly indicating the generation of maximally entangled states.

\begin{figure*}[b]
     \centering
     \includegraphics[width=5.0in]{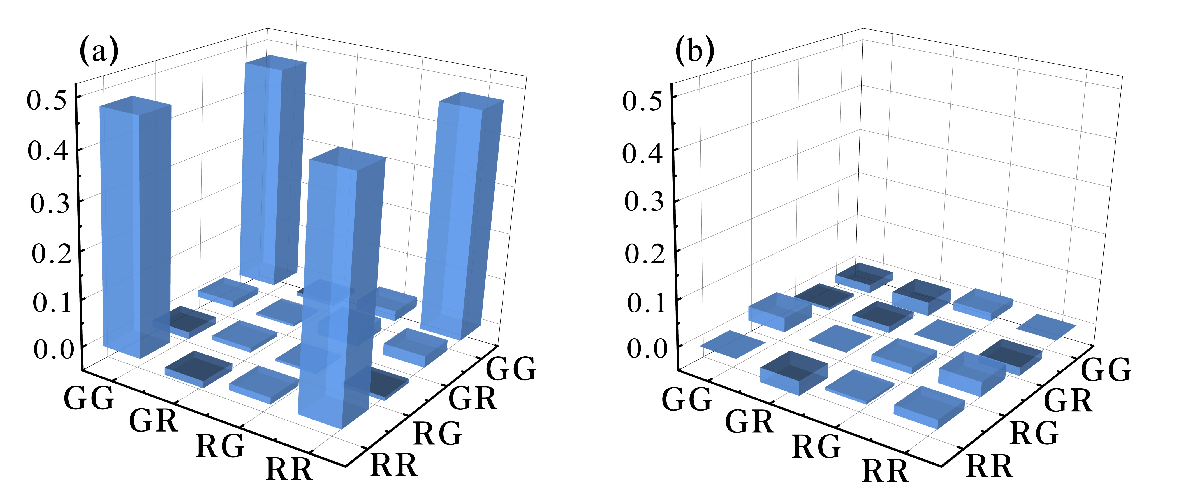}
      \caption{Reconstructed density matrix for OAM entangled state $\mathrm{|\Psi\rangle_{OAM}}$ with the real part (a) and the imaginary part (b). The fidelity of $\mathrm{|\Psi\rangle_{OAM}}$ to the ideal Bell state $|\Psi\rangle=\frac{1}{\sqrt2}         (\left|G_sG_{as}\right\rangle+\left|R_sR_{as}\right\rangle)$ is measured to be 95.7\% and the concurrence is measured to be 0.926.}
            \label{fig:6}
\end{figure*}

The Cauchy-Schwartz inequality, which is defined as 
\begin{equation}
    [\mathrm{g^{2}_{s,as}(\tau)]^2/[g^{2}_{s,s}(0)\cdot{g^{2}_{as,as}(0)}] \le 1}
    \label{eq(5)},
\end{equation}
can be utilized to assess the quantum nature of the biphotons. The maximum value of the autocorrelation functions is $\rm{g^2_{s,s}(0)}$ $=$ $\rm{g^2_{as,as}(0)}=2$, hence the biphotons violate this inequality at least by a factor of 191.8. 

The density matrix of the OAM entangled state, reconstructed through quantum state tomography based on 16 sets of previously obtained coincidence count values, is presented in Fig. \ref{fig:6}. During the reconstruction process, the issue of the density matrix becoming non-positive definite due to experimental errors can be addressed by employing the maximum likelihood estimation algorithm\cite{35,36}. This algorithm helps mitigate the impact of errors and ensures a more accurate reconstruction of the density matrix. Upon obtaining the density matrix, the similarity between the experimentally obtained state and the ideal entangled state can be quantified by calculating the fidelity. The fidelity is defined by
\begin{equation}
    F=Tr(\sqrt{\sqrt{\rho_1}\rho_0\sqrt{\rho_1}})^2.
\end{equation}
Here, $\rho_1$ represents the experimentally obtained density matrix, and $\rho_0$  represents the density matrix of the ideal entangled state (as shown in Eq. \ref{eq(2)}). In our case, the fidelity of the density matrix with respect to the maximum entangled state is measured to be 95.7\%. In addition to fidelity, concurrence ($C$, $0<C\leq1$) is commonly used to quantitatively characterize the quality of an entangled state\cite{37}. A higher concurrence value indicates a state that is closer to maximum entanglement. Based on the reconstructed density matrix, we obtain a concurrence value of 0.926, demonstrating the high quality of the OAM entangled state.
\section{Conclusion}
In conclusion, we propose a simple scheme that utilizes a single driving beam to directly generate OAM entangled biphotons in a hot atomic vapor cell. The measured maximum cross-correlation function value of the biphoton is 27.7, which violates the Cauchy–Schwarz inequality by a factor of 191.8. Additionally, the all-collinear structure enables the driving beam to establish an EIT transparent window for the anti-Stokes photon, extending the coherence time of the biphoton to 40 ns (corresponding to a linewidth of 4 MHz). The OAM entanglement is directly generated by utilizing the OAM conservation law in the SFWM process. Through quantum state tomography, we confirm OAM entanglement of the biphoton with a fidelity of 95.7\% and a concurrence of 0.926 compared to the maximally entangled state. To our knowledge, this represents the first demonstration of generating subnatural-linewidth OAM entangled biphotons in hot atomic systems. 

We note that the entanglement achieved here is limited to two dimensions, as we currently only have the first-order VPP for projection measurements. The potential for achieving higher-dimensional entanglement can be explored by utilizing more versatile measurement devices such as spatial light modulators. Besides, engineering the driving beam mode \cite{38} or selecting perfect optical vortex (POV) modes \cite{39} as measurement basis holds promise for directly generating maximum high-dimensional entanglement. Moreover, the coherence time of the biphoton could be further increased by using a longer Rb cell and placing it inside magnetic shielding cavities.

\section*{Funding.} This work is supported by the National Natural Science Foundation of China (NSFC) (12104358, 12104361, and 12304406) and the Shaanxi Fundamental Science Research Project for Mathematics and Physics (22JSZ004).

\section*{Disclosures.} The authors declare no conflicts of interest.

\section*{Data availability.} Data underlying the results presented in this paper are not publicly available at this time but may be obtained from the authors upon reasonable request.

\bibliography{REF}






\end{document}